\documentclass[aps,prc,twocolumn,preprintnumbers,amsmath,amssymb,superscriptaddress,nofootinbib]{revtex4-2} 

\usepackage{graphicx}
\usepackage{dcolumn}
\usepackage{bm}
\usepackage{color}
\usepackage{threeparttable}
\usepackage{multirow}
\usepackage[dvipsnames]{xcolor}
\usepackage[pdftex,colorlinks,citecolor=blue,urlcolor=blue,linkcolor=blue,bookmarks]{hyperref}
\usepackage{subcaption}

\begin{document}

\title{Ab initio calculation of muon capture on $^{24}$Mg}

\author{L. Jokiniemi}
\affiliation{Departament de Física Quàntica i Astrofísica, Universitat de Barcelona, 08028 Barcelona, Spain}
\affiliation{Institut de Ciències del Cosmos, Universitat de Barcelona, 08028 Barcelona, Spain}
\affiliation{TRIUMF, 4004 Wesbrook Mall, Vancouver, BC V6T 2A3, Canada}

\author{T. Miyagi}%
\affiliation{TRIUMF, 4004 Wesbrook Mall, Vancouver, BC V6T 2A3, Canada}
\affiliation{Technische Universit\"at Darmstadt, Department of Physics, 64289 Darmstadt, Germany}%
\affiliation{ExtreMe Matter Institute EMMI, GSI Helmholtzzentrum f\"ur Schwerionenforschung GmbH, 64291 Darmstadt, Germany}

\author{S. R. Stroberg} 
\affiliation{Department of Physics, University of Washington, Seattle, WA 98195, USA}
\affiliation{Physics Division, Argonne National Laboratory, Lemont IL, 60439, USA}

\author{J. D. Holt}%
\affiliation{TRIUMF, 4004 Wesbrook Mall, Vancouver, BC V6T 2A3, Canada}%
\affiliation{Department of Physics, McGill University, Montr\'eal, QC H3A 2T8, Canada}

\author{J. Kotila}
\affiliation{Finnish Institute for Educational Research, University of Jyv\"askyl\"a, P.O. Box 35, Jyv\"askyl\"a FI-40014, Finland}
\affiliation{Center for Theoretical Physics, Sloane Physics Laboratory, Yale University, New Haven, Connecticut 06520-8120, USA}
\affiliation{%
Department of Physics, University of Jyv\"askyl\"a, P.O. Box 35, Jyv\"askyl\"a FI-40014, Finland
}%

\author{J. Suhonen}
\affiliation{%
Department of Physics, University of Jyv\"askyl\"a, P.O. Box 35, Jyv\"askyl\"a FI-40014, Finland
}%

\date{\today}

\begin{abstract}
In this work we study ordinary muon capture (OMC) on $^{24}$Mg from a first principles perspective. Starting from a particular two- and three-nucleon interaction derived from chiral effective field theory, we use the valence-space in-medium similarity renormalization group (VS-IMSRG) framework to construct effective Hamiltonians and muon-capture operators which nonperturbatively account for many-body physics outside the valence space. The obtained nuclear matrix elements are compared against those from the phenomenological shell model. The impact of including the correlations from the nuclear shell model (NSM) as well as including the induced two-body part is studied in detail. Furthermore, the effects of realistic bound-muon wave function on the operators is studied. Finally, predictions for capture rates to the lowest excited states in $^{24}$Na are given and compared with available data. It is found that the spectroscopic properties of $^{24}$Mg and its OMC daughter $^{24}$Na are fairly well described by both the NSM and VS-IMSRG, and that the effect of the hadronic two-body currents significantly reduces the OMC rates. Both models have some difficulties in matching the measured OMC rates, especially for the $2^+$ final states. This calls for further studies in other light nuclei with available OMC data.
\end{abstract}

\maketitle

\section{\label{sec:intro}Introduction}

Ordinary muon capture (OMC) on nuclei is a nuclear-weak process, in which a negative muon $\mu^-$ is captured by a nucleus $(A,Z)$, resulting in atomic number reduction by one, accompanied by emission of a muon neutrino. 
It can significantly extend the kinematic region of ordinary beta decay, owing to the high energy release and large momentum transfer involved in the process. 
The energy release in this process is on the order of 100 MeV, where the dominant fraction is carried by the neutrino. The large mass of the captured muon facilitates highly forbidden transitions and high excitation energies of the final states. 
These features make muon capture a particularly promising probe for the hypothetical neutrinoless double-beta ($0\nu\beta\beta$) decay \cite{Kortelainen2004,Kortelainen2002}.

Both the $0\nu\beta\beta$ decay and OMC processes involve the axial-vector and pseudoscalar coupling constants $g_{\rm A}$ and $g_{\rm P}$. 
In particular, the half-life of $0\nu\beta\beta$ decay is inversely proportional to $g_{\rm A}^4$. 
However, for decades many theoretical predictions for $\beta$ decays have concluded that one must "quench" the $g_{\rm A}$ coupling in order to reproduce the measured half-lives \cite{Towner1987}. 
While the $g_{\rm A}$ quenching puzzle related to $\beta$ decays was recently solved from first principles in terms of neglected many-body correlations and two-body currents \cite{Gysbers2019}, the possible need for quenching at high-momentum exchange $q\approx 100$ MeV is much less known. 
Since OMC operates at this same momentum-exchange regime, comparing theoretical predictions against measured OMC rates could shed light on this open question. 
Furthermore, while the free proton's pseudoscalar coupling is known to 1\% \cite{Andreev2013}, the correlation effects of $g_{\rm P}$ and corrections to the impulse approximation are under debate. 
In OMC calculations based on the nuclear shell model~\cite{Siiskonen1999,Kortelainen2000,Siiskonen2001}, it has been seen that the Goldberger-Treiman partially conserved axial-vector-current hypothesis $g_{\rm P}/g_{\rm A}\approx 6.8$ is not sufficient to reproduce experimental data.
On the other hand, while the predictions for $g_{\rm P}$ based on chiral perturbation theory ~\cite{Fearing1997,Ando1998,Meissner1998} agree with the value deduced from OMC (within large errors), they disagree with the one required by radiative muon-capture (RMC) experiments.
Hence, OMC serves as an important probe of both these couplings.

The relevance of OMC to $0\nu\beta\beta$ decay is of interest to both experimentalists and theorists \cite{Measday2001}. 
There are several theoretical predictions for the OMC rates and the involved matrix elements based on the nuclear shell model (NSM) \cite{Siiskonen1998,Siiskonen1999,Kortelainen2000,Siiskonen2001,Kortelainen2002,Kortelainen2004,Suzuki2018}, and proton-neutron quasiparticle random-phase approximation (pnQRPA) \cite{Giannaka2015,Jokiniemi2019b,Simkovic2020,Ciccarelli2020} frameworks. More recently, there are also ab initio calculations for the muon-capture rates in very light, $A\leq6$, nuclei based on quantum Monte Carlo methods \cite{Lovato2019,King2021}.
A recent pnQRPA-based study on the OMC strength function in $^{100}$Nb \cite{Jokiniemi2019} showed good agreement with the experimental counterpart measured at RCNP, Osaka
\cite{Hashim2018}. 
Also partial OMC rates to the excited states of intermediate nuclei of several $\beta\beta$-decay triplets have already been measured \cite{Zinatulina2019}, and these studies are planned to be extended to the remaining $\beta\beta$-decay cases.

In particular, OMC on $^{24}$Mg is interesting for many reasons. 
First, the $sd$-shell nuclei $^{24}$Mg and $^{24}$Na are relatively well studied both experimentally and theoretically and accessible to ab initio methods~\cite{Stroberg2016,Novario2020}. 
In addition, the conveniently long lifetime (1067 ns) of the muonic $^{24}$Mg atom  \cite{Suzuki1987}, the well-isolated low-lying $1^+$ states in the final nucleus, $^{24}$Na, and the naturally high abundance of $^{24}$Mg (79 \%) all make it an appealing experimental candidate. 
Partial OMC rates to low-lying states in $^{24}$Na have been measured at TRIUMF \cite{Gorringe1999}, and more recently preliminary measurements aiming to expand the study of OMC on $^{24}$Mg have been performed at PSI, Switzerland \cite{Belov2020}.
In this work we study OMC on $^{24}$Mg from an ab initio perspective, utilizing realistic bound-muon wave functions, for the first time. 
In particular, we use the valence-space in-medium similarity renormalization group (VS-IMSRG)~\cite{Herg16PR,Stroberg2017,Stroberg2019} to  consistently transform Hamiltonians and muon capture operators. 
We compare the obtained nuclear matrix elements against the results computed in the nuclear shell-model framework with the USDB interaction. 
Finally, we compare the resulting capture rates with available experimental data \cite{Gorringe1999} and study the effect of hadronic two-body currents on the computed capture rates. The present work serves a first-step benchmark towards calculating capture rates relevant for all future measurements.

\section{\label{sec:mucap-formalism}Muon-Capture Formalism}

OMC is a semileptonic weak-interaction process similar to electron capture (EC). Here we are interested in the process
\begin{equation}
\label{eq:omc}
\mu^- +\, ^{A}_{Z}{\rm X}(0^+_{\rm g.s.}) \rightarrow \nu_{\mu} +\, ^{~~A}_{Z-1}{\rm Y}(J^{\pi}) \;,
\end{equation}
where a negative muon ($\mu^-$) is captured by the $0^+$ ground state of the even-even nucleus $^{A}_Z{\rm X}$ with atomic number $Z$ and mass number $A$. 
The process leads to the $J^{\pi}$ multipole states of $^{~~A}_{Z-1}{\rm Y}$, the odd-odd isobar of the mother nucleus, of atomic number $Z$; here $J$ is the angular momentum and $\pi$ the parity of the final state. 
At the same time a muon neutrino $\nu_{\mu}$ is emitted. The forbiddenness ($n$) of a muon-capture transition can be defined in the same way as for $\beta$ decay (see Table \ref{tab:forbiddenness}).

In the present study, we compute the corresponding muon capture rates using the formalism derived by Morita and Fujii in 1960 \cite{Morita1960}. This formalism can be translated into another widely used muon-capture theory derived by Foldy, Donnelly and Walecka \cite{Foldy1964,Walecka1975} by introducing multipole operators. However, here we choose to use the Morita-Fujii formalism in which it is straightforward to include realistic bound-muon wave functions, and we  briefly review the formalism in the following subsections.

\subsection{\label{ssec:muonwavefcns}Bound-Muon Wave Functions}

The wave function of a muon bound in an atomic orbit of the mother nucleus can be expressed as an expansion in terms of the normalized spherical spinors $\chi_{\kappa\mu}$ 
\begin{equation}
\psi_{\mu}(\kappa,\mu;\mathbf{r})=\psi_{\kappa\mu}^{(\mu)}=\begin{bmatrix}-iF_{\kappa}\chi_{-\kappa\mu}\\G_{\kappa}\chi_{\kappa\mu}\end{bmatrix}\;,
\label{eq:muonwavefunction}
\end{equation}
where $G_{\kappa}$ and $F_{\kappa}$ are the radial wave functions of the bound state \cite{Morita1960}. Here $\kappa$ denotes the atomic orbit in the following manner
\begin{equation}
\begin{cases}
l=\kappa~\text{and}~j=l-\tfrac12,\quad \text{for}~\kappa>0\\
l=-\kappa-1~\text{and}~j=l+\tfrac12,\quad \text{for}~\kappa<0.
\end{cases}
\end{equation}

After being stopped in the outer shells of an atom, the negative muon goes trough a series of transitions to lower atomic orbitals, leaving it finally on the lowest, $K$ atomic orbit. 
Hence, the captured muon can be assumed to be initially bound in the lowest state $1s_{1/2}$, corresponding to $\kappa=-1$ and $\mu=\pm\tfrac12$. 
Treating the mother nucleus as a point-like particle, we can approximate the wave function of the $1s_{1/2}$ atomic orbit by the Bethe-Salpeter (BS) approximation formula \cite{Bethe1959}. 
Taking $\hbar=c=1$ we then have
\begin{equation}
\begin{split}
G_{-1}&=(2Z/a_0)^{\tfrac32}\sqrt{\frac{1+\gamma}{2\Gamma(2\gamma+1)}}\left(\frac{2Zr}{a_0}\right)^{\gamma-1} e^{-Zr/a_0}\;,\\
F_{-1}&=-\sqrt{\frac{1-\gamma}{1+\gamma}}G_{-1}\;.
\end{split}
\end{equation}
Here $\alpha$ is the fine structure constant, $Z$ the atomic number of the nucleus, $\gamma=\sqrt{1-(\alpha Z)^2}$, and $$a_0=\frac{1}{m'_{\mu}\alpha}$$ is the Bohr radius of the $\mu$-mesonic atom. 
Here we have defined the reduced muon mass as
\begin{equation}
m'_{\mu}=\frac{m_{\mu}}{1+\frac{m_{\mu}}{AM}}
\label{eq:reducedmuonmass}
\end{equation}
where $M$ is the (average) nucleon mass, and $AM$ is the mass of the mother (and daughter) nucleus. 
For a light nucleus, as in the present case, $\alpha Z$ is small and we can approximate $\gamma\approx 1$, leading to
\begin{equation}
\begin{split}
G_{-1}&=2(\alpha Zm'_{\mu})^{\tfrac32}e^{-\alpha Z m'_{\mu}r}\;,\\
F_{-1}&=0\;.
\end{split}
\label{eq:pointlikeapproximation}
\end{equation}
The approximation is similar to the hydrogen-like Schr\"odinger equation with a modified $a_0$. 
This is the adopted form for the bound-muon wave function in Ref.~\cite{Morita1960}, and also our starting point in the present study. 

\begin{table}
\caption{Forbiddenness rules for odrinary muon capture.}
\begin{ruledtabular}
    \centering
    \begin{tabular}{ccc}
          $n$ &Spin change $\vert J_f-J_i \vert$ &Parity change $\pi_f\pi_i$\\
         \hline
         0 & 0,1 & $+1$\\
         1 & 0,1,2 &$-1$\\
         $\geq2$ &$n$, $n+1$ &$(-1)^n$
    \end{tabular}
    \label{tab:forbiddenness}
    \end{ruledtabular}
\end{table}

In order to take the finite size of the nucleus properly into account, we construct a realistic bound-muon wave function by solving the Dirac wave equations~\cite{Jokiniemi2021} for the large, $G_{-1}$, and small ,$F_{-1}$, parts of the wave function \eqref{eq:muonwavefunction} in the Coulomb field created by the nucleus.
Assuming the muon is bound in the lowest state $1s_{1/2}$ ($\kappa=-1$), the components satisfy the coupled differential equations
\begin{equation}
\begin{cases}
\frac{\rm d}{{\rm d}r}G_{-1}+\frac{1}{r}G_{-1}=\frac{1}{\hbar c}(mc^2-E+V(r))F_{-1}\;,\\
\frac{\rm d}{{\rm d}r}F_{-1}-\frac{1}{r}F_{-1}=\frac{1}{\hbar c}(mc^2+E-V(r))G_{-1}\;.
\end{cases}
\label{eq:coupled-differentials}
\end{equation}

Taking a uniform distribution of the nuclear charge within the charge radius $R_c=r_0A^{1/3}$, the potential energy $V(r)$ in Eqs.~\eqref{eq:coupled-differentials} can be written:
\begin{equation}
V(r)=
  \begin{cases}
-\frac{Ze^2}{2R_c}\left[3-\left(\frac{r}{R_c}\right)^2\right]\;, & \text{if $r\leq R_c$} \\
-\frac{Ze^2}{r}\;, & \text{if $r>R_c$.}
  \end{cases}
\label{eq:muonpotentialenergy}
\end{equation}
These equations \eqref{eq:coupled-differentials} can then be solved by means of the package \textsc{Radial} \cite{Salvat1995}
using a piece-wise-exact power-series expansion of the radial functions, which are summed to the prescribed accuracy. 
A similar method has previously been used for bound-electron wave functions in the context of $\beta\beta$ decay \cite{Kotila2012,Kotila2013}.

\begin{figure}[h!]
    \centering
    \includegraphics[width=0.9\linewidth]{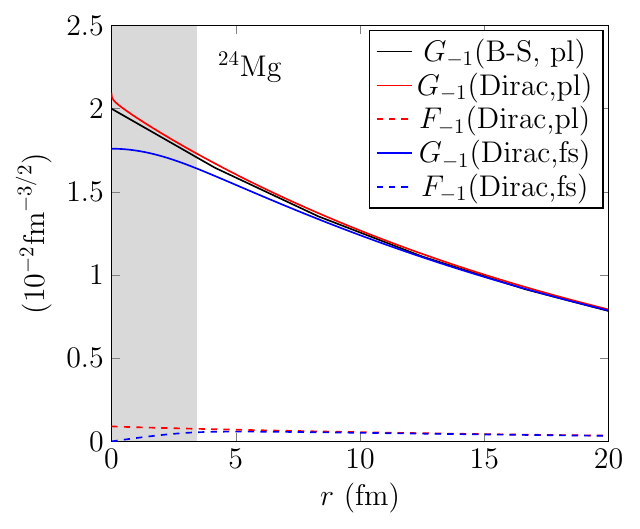}
    \caption{Large ($G_{-1}$) and small ($F_{-1}$) components of the bound-muon wave functions in $^{24}$Mg. 'B-S' refers to the BS approximation and 'Dirac' to the exact wave function from the Dirac equation. 'pl' and 'fs' refer to point-like nucleus and finite-size nucleus. The shaded area corresponds to the interior of the nucleus.}
    \label{fig:muonwavefunctions}
\end{figure}

In Fig.~\ref{fig:muonwavefunctions} we plot the amplitudes $G_{-1}$ and $F_{-1}$ as solved from the Dirac equations (solid and dashed blue lines, correspondingly) and contrast them with those obtained from the BS approximation (black line) of 
Eq. \eqref{eq:pointlikeapproximation}. 
The small part vanishes in the BS approximation and thus does not appear in the figure. 
For comparison, we have also plotted the exact solution of the Dirac equation corresponding to point-like nucleus (solid and dashed red lines). 
The behavior of the Dirac wave function corresponding to the point-like nucleus is essentially similar to the BS approximation. 
When the finite size of the nucleus is taken into account, the Dirac wave function is notably suppressed at $r\lesssim 5$ fm, i.e. inside the nucleus.
As illustrated in the figure, the small part of the exact solution is negligible compared to the large part, hence we can safely neglect it in the calculations. 
This is expected, since the small part is suppressed by $v/c=Z\alpha$.

{ 
\renewcommand{\arraystretch}{2.0}
\begin{table*}[t]\centering
\caption{Definition of $\mathcal{O}_{s}$ in Eq. \eqref{eq:ME-definition} for different OMC nuclear matrix elements (NMEs).
}
\begin{ruledtabular} 
\begin{tabular}{ll}
\centering NME & $\mathcal{O}_s$\\
\hline
$\mathcal{M}[0\,w\,u]$ &$j_w(qr_s)G_{-1}(r_s)\mathcal{Y}_{0wu}^{M_f-M_i}(\hat{\mathbf{r}}_s)\delta_{wu}$\\
$\mathcal{M}[1\,w\,u]$ &$j_w(qr_s)G_{-1}(r_s)
\mathcal{Y}_{1wu}^{M_f-M_i}(\hat{\mathbf{r}}_s,\boldsymbol{\sigma}_s)$\\
$\mathcal{M}[0\,w\,u\,\pm]$ &$[j_w(qr_s)G_{-1}(r_s)\mp\frac{1}{q} j_{w\mp 1}(qr_s)\frac{d}{dr_s}G_{-1}(r_s)]
\mathcal{Y}_{0wu}^{M_f-M_i}(\hat{\mathbf{r}}_s)\delta_{wu}$\\
$\mathcal{M}[1\,w\,u\,\pm]$ &$[j_w(qr_s)G_{-1}(r_s)\mp\frac{1}{q} j_{w\mp 1}(qr_s)\frac{d}{dr_s}G_{-1}(r_s)]
\mathcal{Y}_{1wu}^{M_f-M_i}(\hat{\mathbf{r}}_s,\boldsymbol{\sigma}_s)$\\
$\mathcal{M}[0\,w\,u\,p]$ &$ij_w(qr_s)G_{-1}(r_s)\mathcal{Y}_{0wu}^{M_f-M_i}(\hat{\mathbf{r}}_s)
\boldsymbol{\sigma}_s\cdot \mathbf{p}_s \delta_{wu}$\\
$\mathcal{M}[1\,w\,u\,p]$ &$ij_w(qr_s)G_{-1}(r_s)
\mathcal{Y}_{1wu}^{M_f-M_i}(\hat{\mathbf{r}}_s,\mathbf{p}_s)$\\
\end{tabular}
\end{ruledtabular}
\label{tab:operators}
\end{table*}
}

\subsection{Muon-Capture Matrix Elements}
\label{ssec:OMC-MEs}

We compute the OMC matrix elements using the formalism originally developed by Morita and Fujii in \cite{Morita1960}. 
The formalism takes into account both the genuine and induced vector and axial-vector weak nucleon currents. 
The formalism is rather involved, and here we only present the main ingredients needed in the calculations.
Further details on the derivations of the equations can be found in \cite{Morita1960,Jokiniemi-phd}.
Here it is appropriate to note that in Ref.~\cite{Menendez2011} it was found that momentum-dependent two-body hadronic currents could have an impact on the $0\nu\beta\beta$-decay nuclear matrix elements (NMEs). Owing to the similar momentum-exchange scales, the effects of two-body currents could be expected to be similar for OMC. The currents have already been included in OMC calculations of light nuclei \cite{Lovato2019,King2021}, but the effects in heavier nuclei have yet to be explored. In the present work, we study the effect of the two-body currents by including approximate normal-ordered two-body corrections derived in \cite{Hoferichter2020} for the axial-vector and pseudoscalar currents. Furthermore, in Refs.~\cite{Lovato2019,King2021} it was shown that the vector two-body current also has an impact on the capture rates; we will explore this in future work.

The matrix elements for a transition from an initial $J_i$ state to a final $J_f$ state can be defined as
\begin{equation}
\begin{split}
&\int \Psi_{J_fM_f}\sum_{s=1}^{A}\mathcal{O}_s\tau^s_-\Psi_{J_iM_i}
d\mathbf{r}_1...d\mathbf{r}_A\\
&=\mathcal{M}[k\,w\,u\,\binom{\pm}{p}](J_i\ M_i\ u\ M_f-M_i|J_f\,M_f)\;,
\end{split}
\label{eq:ME-definition}
\end{equation}
where $\Psi_{J_iM_i}$ and $\Psi_{J_fM_f}$ are the nuclear wave functions corresponding to the initial and final state. 
The operators $\mathcal{O}_{s}$ in \eqref{eq:ME-definition} are defined in Table \ref{tab:operators}. 
Here we assume that the muon is bound on the $\kappa=-1$ orbit and that the small component of the bound-muon wave function is negligible as clearly demonstrated in Fig.~\ref{fig:muonwavefunctions}. 
The small component is only about 0.1\% of the large component at $r<R_c$.
Neglecting the small component simplifies the expressions of the matrix elements considerably (see Table I of Ref.~\cite{Morita1960}).

In Table~\ref{tab:operators}, $j_w(qr_s)$ is the spherical Bessel function of rank $w$. 
The quantities $\mathcal{Y}_{kwu}^M$ are the (vector) spherical harmonics defined as
\begin{equation}
\begin{split}
\mathcal{Y}_{0wu}^M(\hat{\mathbf{r}})\equiv & (4\pi)^{-1/2}Y_{w,M}(\hat{\mathbf{r}})\;,\\
\mathcal{Y}_{1wu}^M(\hat{\mathbf{r}},\boldsymbol{\sigma})\equiv
&\sum_m(1\ -m\ w\ m+M|u\,M)\\
&\times Y_{w,m+M}(\hat{\mathbf{r}})\sqrt{\frac{3}{4\pi}}\sigma_{-m}\;,\\
\mathcal{Y}_{1wu}^M(\hat{\mathbf{r}},\mathbf{p})\equiv
&\sum_m(1\ -m\ w\ m+M|u\,M)\\
&\times Y_{w,m+M}(\hat{\mathbf{r}})\sqrt{\frac{3}{4\pi}}p_{-m}\;,\\
\end{split}
\label{eq:spherical-harmonics}
\end{equation}
where $\boldsymbol{\sigma}$ is the Pauli spin vector, $\mathbf{p}$ is the nucleon momentum, $Y_{w,M}(\hat{\mathbf{r}})$ are the spherical harmonics, and $\hat{\mathbf{r}}$ is the unit coordinate vector for angles in spherical coordinates.

The $q$ in Table~\ref{tab:operators} is the $Q$-value of the capture process:
\begin{equation}
q=(m_{\mu}-W_0)\left(1-\frac{m_{\mu}}{2(m_{\mu}+AM)}\right)\;,
    \label{eq:q}
\end{equation}
where $W_0=M_f-M_i+m_e+E_X$. 
Here $M_f$($M_i$) is the nuclear mass of the final(initial) nucleus, $m_e$ the rest mass of an electron, $m_{\mu}$ the rest mass of a muon, $M$ the average nucleon mass, and $E_X$ the excitation energy of the final $J^{\pi}$ state. 
At low excitation energies, $W_{0}/m_{\mu} \ll 1$, so the operators do not significantly depend on the excitation energy.

In NSM calculations, the matrix elements of Eq.~\eqref{eq:ME-definition} are expressed in terms of reduced matrix elements:
\begin{equation}
\begin{split}
\mathcal{M}_{\mu}=&\widehat{J}_{f}^{-1}\sum_{pn}
\mathcal{O}_{\mu,np}
T^{u}_{J_{f}J_{i},np},
\end{split}
\end{equation}
where indices $p$ and $n$ label proton and neutron orbitals, respectively.
Here $T^{u}_{J_{f}J_{i},np}$ is the one-body transition density (OBTD) 
\begin{equation}
\begin{aligned}
T^{u}_{J_{f}J_{i},np} &= \frac{1}{\hat{u}} \langle J_{f} || [c^{\dag}_{n}\tilde{c}_{p}]_{u}|| J_{i} \rangle, \\
\tilde{c}_{p} &= (-1)^{j_{p}-m_{p}} c_{p}
\end{aligned}
\end{equation}
and we adopt the shorthand notation $\mu = \left[kwu\binom{\pm}{p}\right]$ for one-body transition matrix elements $\mathcal{O}_{\mu,np}$, corresponding to the operators given in Table \ref{tab:operators}. 
We evaluate these matrix elements in the harmonic oscillator (HO) basis.

\subsection{\label{ssec:rates}Capture Rates}

The capture rate for a transition from a $J_i^{\pi}$ initial state to a $J_f^{\pi}$ final state can be written as
\begin{equation}
  W=2P\frac{2J_f+1}{2J_i+1}\left(1-\frac{q}{m_{\mu}+AM}\right)q^2 \;,
\end{equation}
with 
\begin{equation}
\label{eq:p}
P=\frac{1}{2}\sum_{\kappa u}\Big|A+B+C+D+E+F+G\Big|^2\;,
\end{equation}
where the quantities $A$-$G$ depend on $\kappa$ and $u$ and are defined as
\begin{equation}
\label{eq:A}
A=g_{\rm V}\mathcal{M}[0\,l\,u]S_{0u}(\kappa)\delta_{lu}\;,
\end{equation}
\begin{equation}
\label{eq:B}
B=g_{\rm A}\mathcal{M}[1\,l\,u]S_{1u}(\kappa)\;,
\end{equation}
\begin{equation}
\label{eq:C}   
C=-\frac{g_{\rm V}}{M}
\mathcal{M}[1\,\bar{l}\,u\,p]S'_{1u}(-\kappa)\;,
\end{equation}
\begin{equation}
\label{eq:D}
\begin{split}
D=&\sqrt{3}\frac{g_{\rm V}q}{2M}\bigg(\sqrt{\frac{\bar{l}+1}{2\bar{l}+3}}
\mathcal{M}[0\,\bar{l}\!+\!1\,u\,+]\delta_{\bar{l}+1,u}\\
 &+ \sqrt{\frac{\bar{l}}{2\bar{l}-1}}\mathcal{M}[0\,\bar{l}\!-\!1\,u\,-]\delta_{\bar{l}-1,u}\bigg)S'_{1u}(-\kappa)\;,
\end{split}
\end{equation}
\begin{equation}
\label{eq:E}  
\begin{split}
E=&\sqrt{\frac{3}{2}}\left(\frac{g_{\rm V}q}{M}\right)(1+\mu_p-\mu_n)S'_{1u}(-\kappa)\\
&\times\Big(\sqrt{\bar{l}+1}W(1\,1\,u\,\bar{l}\,;1\,\bar{l}+1)
\mathcal{M}[1\,\bar{l}\!+\!1\,u\,+]\\
 &+\sqrt{\bar{l}}W(1\,1\,u\,\bar{l}\,;1\,\bar{l}-1)\mathcal{M}[1\,\bar{l}\!-\!1\,u\,-]\Big)\;,
\end{split}
\end{equation}
\begin{equation}
\label{eq:F} 
F=-\left(\frac{g_{\rm A}}{M}\right)\mathcal{M}[0\,\bar{l}\,u\,p]S'_{0u}(-\kappa)
\delta_{\bar{l}u}\;,
\end{equation}
and
\begin{equation}
\label{eq:G} 
\begin{split}
G=&\sqrt{\frac{1}{3}}\frac{q}{2M}(g_{\rm P}-g_{\rm A}) S'_{0u}(-\kappa)\delta_{\bar{l}u}\\
&\times\bigg(\sqrt{\frac{\bar{l}+1}{2\bar{l}+1}}\mathcal{M}[1\,\bar{l}\!+\!1\,u\,+]+
\sqrt{\frac{\bar{l}}{2\bar{l}+1}}\mathcal{M}[1\,\bar{l}\!-\!1\,u\,-]\bigg)\;. 
\end{split}
\end{equation}
The $W(...)$ in Eqs. (\ref{eq:A}-\ref{eq:G}) are the usual Racah coefficients and the $S$'s are geometric factors defined as
\begin{equation}
S_{ku}(\kappa)=
\begin{cases}
\sqrt{2(2j+1)}W(\tfrac12\, 1\,j\,l\,;\tfrac12 \,u)\delta_{lw}\text{ , for } k=1\\
\sqrt{\frac{2j+1}{2l+1}}\delta_{lw}\text{ , for } k=0
\end{cases}
\end{equation}
and
\begin{equation}
S'_{ku}(-\kappa)={\rm sgn}(\kappa)S_{ku}(-\kappa)\;,
\end{equation}
where ${\rm sgn}(\kappa)$ is the sign of $\kappa$.
The angular momenta $l$ and $\bar{l}$ correspond to $\kappa$ and $-\kappa$ respectively. 
We use the Goldberger-Treiman partially conserved axial-vector-current (PCAC) value \begin{equation}
g_{\rm P}/g_{\rm A}=\frac{2Mq}{q^2+m_{\pi}^2}\approx 6.8
\end{equation}
for the ratio of the pseudoscalar and axial-vector coupling strengths. Note that in this formalism, the pseudoscalar interaction is written as $(g_{\rm P}/2M)\boldsymbol{\sigma}\cdot\mathbf{p}$.
For the axial-vector coupling we use the free-nucleon value $g_{\rm A}=1.27$. 
The explicit expressions for the $P$ of Eq. \eqref{eq:p} containing all the next-to-leading order terms can be found e.g. in Ref.~\cite{Jokiniemi-phd}. 
Note that in Ref.~\cite{Morita1960} the terms of the order $1/M^2$ were omitted from the explicit expressions for $P$.

\subsection{\label{ssec:twobodycurrents}Two-Body Currents}

\begin{figure}[h!]
    \centering
    \includegraphics[width=\linewidth]{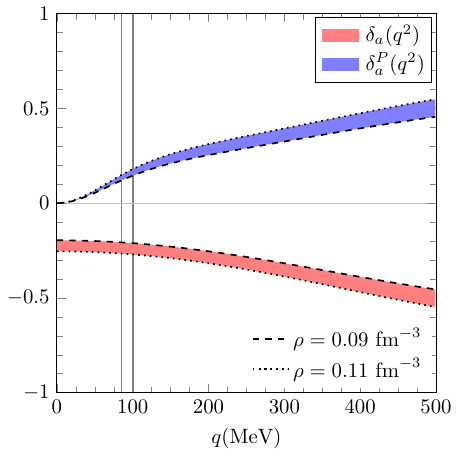}
    \caption{Two-body currents used in the present work as functions of momentum-exchange. The dashed lines denote the currents obtained by $\rho=0.09~{\rm fm}^{-3}$ and the dotted those obtained with $\rho=0.11~{\rm fm}^{-3}$. The typical momentum exchange region of the transitions considered in the present work is denoted by vertical gray lines.}
    \label{fig:two-body-currents}
\end{figure}

We take the effect of two-body currents (2BCs) into account by replacing 
$$g_{\rm A}\rightarrow (1+\delta_a(q^2))g_{\rm A}$$ and $$g_{\rm P}\rightarrow (1-\frac{q^2+m_{\pi}^2}{q^2}\delta_a^P(q^2))g_{\rm P}\;,$$
where $\delta_a(q^2)$ and $\delta_a^P(q^2)$ are approximated by the normal-ordered one-body part of two-body currents with respect to a Fermi-gas reference state with density $\rho$ \cite{Hoferichter2020} as
\begin{equation}
\begin{split}
\delta_a(q^2)=&-\frac{\rho}{F_{\pi}^2}\bigg[\frac{1}{3}\left(c_4+\frac{1}{4M}\right)[3I^{\sigma}_2(\rho,q)-I^{\sigma}_1(\rho,q)]\\
&-\frac13\bigg(c_3-\frac{1}{4M}\bigg)I^{\sigma}_1(\rho,q)\\
&-\frac{1}{12}\left(c_6+\frac{1}{4M}\right)I_{c6}(\rho,q)-\frac{c_D}{4g_A\Lambda_{\chi}}\bigg]\;,
\end{split}
    \label{eq:delta_a}
\end{equation}
\begin{equation}
\begin{split}
\delta_a^P(q^2)=&\frac{\rho}{F_{\pi}^2}\bigg[-2(c_3-2c_1)\frac{M_{\pi}^2q^2}{(M_{\pi}^2+q^2)^2}\\
&+\frac13(c_3+c_4)I^P(\rho,q)\\
&-\bigg(\frac{1}{12}\Big(c_6+\frac{1}{4M}\Big)-\frac23\frac{c_1M_{\pi}^2}{M_{\pi}^2+q^2}\bigg)I_{c6}(\rho,q)\\
&-\frac{q^2}{M_{\pi}^2+q^2}\bigg(\frac{c_3}{3}[I^{\sigma}_1(\rho,q)+I^P(\rho,q)]\\
&+\frac{1}{3}\Big(c_4+\frac{1}{4M}\Big)[I^{\sigma}_1(\rho,q)+I^P(\rho,q)\\&-3I^{\sigma}_2(\rho,q)]\bigg)
-\frac{c_D}{4g_A\Lambda_{\chi}}\frac{q^2}{M_{\pi}^2+q^2}\bigg]\;.
\end{split}
    \label{eq:delta_a^P}
\end{equation}
We take the integrals $I^{\sigma}_1(\rho,q)$, $I^{\sigma}_2(\rho,q)$, $I_{c6}(\rho,q)$ and $I^P(\rho,q)$ from \cite{Klos2015} and use the density range $\rho=0.09\dots 0.11~{\rm fm}^{-3}$. We use the same constants as in \cite{Hoferichter2020}: $M_{\pi}=138.04~{\rm MeV}/c^2$, $F_{\pi}=92.28~{\rm MeV}/c^2$ and $\Lambda_{\chi}=700~{\rm MeV}$. We follow the approach of \cite{Hu2022} and take the low-energy constants (LECs) $c_1,~c_2,c_4$ and $c_D$ to be the same as in the 1.8/2.0(EM) interaction that is used in the IMSRG calculations. Note that the relativistic $1/M$ corrections are absorbed in the chosen parameters in the previous NSM calculations \cite{Hoferichter2020}, while here they are explicitly taken into account by replacing $c_4\rightarrow c_4+1/(4M)$ and $c_6\rightarrow c_6+1/(4M)$. We list the chosen LECs in Table \ref{tab:LECs}. The corresponding two-body corrections are depicted in Fig.~\ref{fig:two-body-currents}, where the relevant momentum-exchange region is denoted by vertical gray lines. The values of $\delta_a(q^2)$ and $\delta_a^P(q^2)$ for OMC to the different final states considered in the present study are listed in Table \ref{tab:2BCs} in the Appendix. At zero-momentum transfer, relevant for $\beta$ decays, the axial-vector two-body current corresponds to quenching of $g_{\rm A}$ by a factor $\approx 0.73-0.70$.

\begin{table}[ht!]
    \centering
    \caption{Low-energy constants corresponding to the EM18./2.0 interaction \cite{Hebeler2011,Simonis2017}. The $c_i$ coefficients have units ${\rm GeV}^{-1}$ while $c_D$ is dimensionless.}
    \begin{ruledtabular}
    \begin{tabular}{ccD{.}{.}{5}c}
    & LEC &\multicolumn{1}{c}{EM1.8/2.0}\\
    \hline
         & $c_1$ & -0.81\\ 
         & $c_3$ & -3.20\\
         & $c_4$ &  5.40\\
         & $c_6$ &  5.01\\
         & $c_D$ &  1.264\\
    \end{tabular}
    \end{ruledtabular}
    \label{tab:LECs}
\end{table}

\section{\label{sec:many-body-methods}Many-Body Methods}

\subsection{\label{ssec:imsrg}Valence-space in-medium similarity renormalization group}
In this work we use the ab initio VS-IMSRG~\cite{Tsukiyama2012,Bogner2014,Stroberg2019,Miya20lMS}, to constructs a continuous ($s$-dependent) unitary transformation $U(s)$, to decouple an effective valence-space Hamiltonian $U(s)HU^{\dag}(s)$ from the full Hilbert space.
In the VS-IMSRG, based on the exponential ansatz $U(s)=e^{\Omega(s)}$ with the anti-Hermitian operator $\Omega(s)$, one finds $\Omega(s)$ by solving the flow equation~\cite{Morris2015}:
\begin{equation}
\label{eq:flow-eq}
\frac{d\Omega}{ds} = \sum_{n=1}^{\infty} \frac{B_{n}}{n!} {\rm ad}^{n}_{\Omega}(\eta), \quad
{\rm ad}^{n}_{\Omega} = [\Omega, {\rm ad}^{n-1}_{\Omega}],
\end{equation}
with ${\rm ad}^{0}_{\Omega} = \eta$ and $\Omega(0)=0$. 
The object $\eta$ is known as the generator of the flow equation, and we use the arctangent generator in this work.
With the same unitary transformation, any valence-space effective operator can be derived consistently \cite{Parz17Trans}.

The actual calculation procedure is as follows. 
We begin from a nuclear Hamiltonian based on chiral effective field theory \cite{Epel09RMP,Mach11PR} expressed in the 13 major-shell HO space at the frequency 16~MeV. 
In the current study, the employed interaction is the two (NN) and three-nucleon (3N) force 1.8/2.0 (EM)~\cite{Hebeler2011,Simonis2017}, where the NN force is given at order N$^{3}$LO and the 3N force at N$^{2}$LO.
This interaction has been shown to reproduce ground-state energies globally to the $^{132}$Sn region and beyond~\cite{Morr18Tin,Stroberg2021,Miya21Heavy}, while generally giving too small radii \cite{Groo20Cu}.
For the 3N piece, due to memory limitations, we need to introduce an additional truncation $E_{\rm 3max}=24$~\cite{Miya21Heavy} defined as the sum of the three-body HO quanta.
Before solving the flow equation~\eqref{eq:flow-eq}, we optimize the single-particle orbitals through transformation to the Hartree-Fock (HF) basis.
For OMC (or $\beta$ decays), two possible HF reference states can be considered, either from the parent or daughter nucleus. In the absence of any approximation, the result of the many-body calculation should be independent of the choice of the reference state, so the reference-state sensitivity can be used as one tool to gauge the error of the many-body approximation.
Since the evolution of the full 3N Hamiltonian in a realistic model space is challenging~\cite{Heinz2021}, we employ the ensemble normal-ordering~\cite{Stroberg2017} technique to capture 3N forces between valence nucleons. 

During the VS-IMSRG evolution, all the operators are truncated at the two-body level, referred to as the VS-IMSRG(2) approximation.
As a consequence, the originally one-body OMC operators have both one- and induced two-body terms:
$$
\overline{\mathcal{O}}_{\mu} 
= e^{\Omega(s)} \mathcal{O}_{\mu}
e^{-\Omega(s)}
\approx \overline{\mathcal{O}}_{\mu}^{(1)} + \overline{\mathcal{O}}_{\mu}^{(2)}
$$
Combining the consistently evolved operators and the one- and two-body transition density matrices, we compute the NMEs:
\begin{equation}
\begin{aligned}
\hat{J}_{f}\mathcal{M}_{\mu} 
\approx &\sum_{pq} \overline{\mathcal{O}}^{(1)}_{\mu, pq}
T^{u}_{J_{f}J_{i},pq} \\
&+ \frac{1}{4} \sum_{pqrs}\sum_{J_{pq}J_{rs}} \overline{\mathcal{O}}^{(2) J_{pq}J_{rs}}_{\mu,pqrs}
T^{J_{pq}J_{rs} u}_{J_{f}J_{i}, pqrs},
\end{aligned}
\end{equation}
with the evolved OMC operator matrix elements $\overline{O}^{(1)}_{\mu,pq}$ and $\overline{O}^{(2)J_{pq}J_{rs}}_{\mu,pqrs}$,
and $p$, $q$, $r$ and $s$ running through all possible proton and neutron states.
In addition to the OBTD $T^{u}_{J_{f}J_{i},pq}$, the two-body transition density $T^{J_{pq}J_{rs} u}_{J_{f}J_{i}, pqrs}$ is introduced as
\begin{equation}
\begin{aligned}
T^{J_{pq}J_{rs} u}_{J_{f}J_{i}, pqrs} &= \frac{1}{\hat{u}}
\langle J_{f} || [A_{pq,J_{pq}}^{\dag} \tilde{A}_{rs,J_{rs}}]_{u}|| J_{i} \rangle, \\
A^{\dag }_{pq,J_{pq}M_{pq}} &= \frac{1}{\sqrt{1+\delta_{pq}}} [c^{\dag}_{p} c^{\dag}_{q}]_{J_{pq}M_{pq}}, \\
\tilde{A}_{pq,J_{pq}M_{pq}} &= \frac{1}{\sqrt{1+\delta_{{p}{q}}}} [\tilde{c}_{p} \tilde{c}_{q}]_{J_{pq}-M_{pq}}.
\end{aligned}
\end{equation}
Note that $\delta_{pq}$ indicates $\delta_{n_{p}n_{q}}\delta_{l_{p}l_{q}}\delta_{j_{p}j_{q}}\delta_{t_{z,p}t_{z,q}}$ with the nodal quantum number, orbital angular momentum, total angular momentum, and label distinguishing proton and neutron, respectively.
The flow equations are solved with \texttt{imsrg++} code~\cite{imsrg++}, and the valence-space diagonalization and computation of the corresponding transition densities are performed with the \texttt{KSHELL} code~\cite{Shimizu2019}.

\subsection{\label{ssec:structure}Nuclear Shell Model}

\begin{figure*}[h!]
\includegraphics[width=\linewidth]{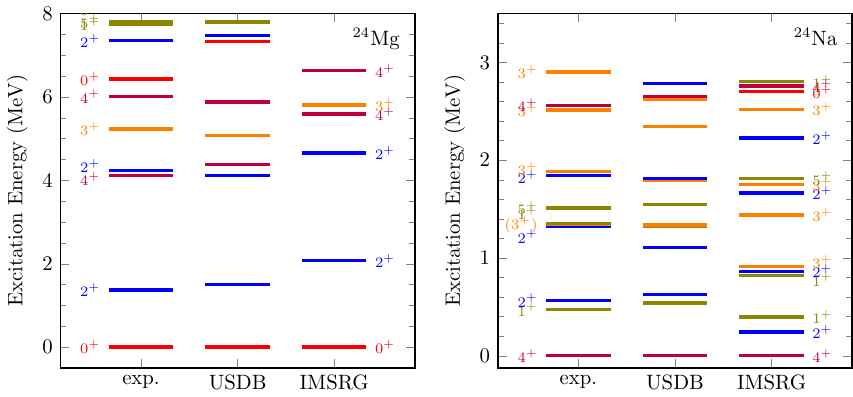}
\caption{Excitation energies in $^{24}$Mg and $^{24}$Na. Only the positive-parity states are show. For the cases, where the levels are very close to each other, they have been slightly shifted for better visibility.}
\label{fig:excitation_energies}
\end{figure*}

\begin{table*}[ht!]
    \centering
        \caption{Energies, magnetic dipole and electric quadrupole moments of excited states in $^{24}$Mg and $^{24}$Na. Experimental values are taken from \cite{NDS}.}
        \begin{ruledtabular}
    \begin{tabular}{ccccccccccc}
     Nucleus &$J_i^{\pi}$ &\multicolumn{3}{c}{$E({\rm MeV})$}&\multicolumn{3}{c}{$\mu(\mu_{\rm N})$}&\multicolumn{3}{c}{$Q(e^2\rm fm^2)$}\\
    \cline{3-5} \cline{6-8} \cline{9-11}
            &                     &exp. &NSM &IMSRG &exp. &NSM &IMSRG &exp. &NSM &IMSRG\\
            \hline  
    $^{24}$Mg &$2^+$ &1.369 & 1.502 &1.981 &1.08(3) &1.008 &1.033 &$-29(3)$ &$-19.346$ &$-12.9$\\
    $^{24}$Mg &$4^+$ &4.123 & 4.372 &5.327 &1.7(12) &2.021 &2.096 &-\\
    $^{24}$Mg &$2^+$ &4.238 & 4.116 &4.327 &1.3(4)  &1.011 &1.085 &-\\
    $^{24}$Mg &$4^+$ &6.010 & 5.882 &6.347 &2.1(16) &2.015 &2.089 &-\\
    $^{24}$Na &$4^+$ &0.0   & 0.0 &0.0   &1.6903(8) &1.533 &1.485 &-\\
    $^{24}$Na &$1^+$ &0.472 & 0.540 &0.397 &$-1.931(3)$ &$-1.385$ &$-0.344$ &-\\
    \end{tabular}
    \end{ruledtabular}
    \label{tab:electromagnetic_moments}
\end{table*}

\begin{table*}[ht!]
    \centering
        \caption{Experimental and calculated E2 and M1 decays of $^{24}$Mg and $^{24}$Na. Experimental values are derived from \cite{NNDC}. The dashes indicate missing experimental data.}
        \begin{ruledtabular}
    \begin{tabular}{cccccccc}
    Nucleus &$J_i\rightarrow J_f$ &\multicolumn{3}{c}{$B(E2)(e^2{\rm fm}^4)$} &\multicolumn{3}{c}{$B(M1)(\mu_{\rm N}^2)$}\\
    \cline{3-5}\cline{6-8}
            &                     &exp. &NSM &IMSRG &exp. &NSM &IMSRG\\
            \hline
    $^{24}$Mg &$2^+_1\rightarrow 0^+_{\rm gs}$ &$88\pm 4$ &95 &48 &- & & \\
    $^{24}$Mg &$4^+_1\rightarrow 2^+_1$ &$160\pm 20$ &125 &63 &- & & \\
    $^{24}$Mg &$2^+_2\rightarrow 2^+_1$ &$15\pm 2$ &19 &14 &- & & \\
    $^{24}$Mg &$2^+_2\rightarrow 0^+_{\rm gs}$ &$8.0\pm 0.8$ &8.9 &2.8 &- & & \\
    $^{24}$Mg &$3^+_1\rightarrow 2^+_2$ &$240\pm 30$ &170 &84 &- & & \\
    $^{24}$Mg &$3^+_1\rightarrow 2^+_1$ &$10\pm 2$ &14 &4 &- & & \\
    $^{24}$Mg &$0^+_2\rightarrow 2^+_2$ &$37\pm 6$ &18 &5 &- & & \\
    $^{24}$Mg &$0^+_2\rightarrow 2^+_1$ &$2.6\pm 0.4$ &0.2 &0.5 &- & & \\
    $^{24}$Mg &$2^+_3\rightarrow 0^+_{\rm gs}$ &$2.8\pm 0.9$ &0.6 &0.2 &- & & \\
    $^{24}$Na &$3^+_1\rightarrow 4^+_{\rm gs}$ &- & & &$0.34\pm 0.09$ &0.18 &0.44 \\
    $^{24}$Na &$2^+_3\rightarrow 1^+_1$ &$5\pm 4$ &2 &3 &$0.20\pm 0.03$ &0.005 &0.27 \\
    $^{24}$Na &$3^+_2\rightarrow 2^+_1$ &($1\substack{+3 \\ -1}$) &3 &2 &$0.41\pm 0.09$ &0.52 &0.07 \\
    $^{24}$Na &$3^+_2\rightarrow 4^+_{\rm gs}$ &$1.7\pm 1.0$ &1.6 &0.4 &$0.09\pm 0.02$ &0.08 &0.03 \\
    \end{tabular}
    \end{ruledtabular}
    \label{tab:magnetic_decays}
\end{table*}

\begin{table}[ht!]
    \centering
    \begin{threeparttable}
        \caption{Experimental and calculated $\log ft$ values for $\beta$ decays of $^{24}$Na leading to excited states in $^{24}$Mg, respectively. Experimental values are taken from \cite{NNDC}. The results include the typical quenching of $0.77\pm 0.02$ for the NSM and the effect of axial-vector two-body current at $|\mathbf{q}|=0$ for the IMSRG. In the case of IMSRG, an additional variation comes from the reference-state dependence.}
        \begin{ruledtabular}
    \begin{tabular}{ccccc}
    Nucleus &$J_i\rightarrow J_f$ &\multicolumn{3}{c}{$\log ft$}\\
    \cline{3-5}
            &                     &exp. &NSM &IMSRG\\
            \hline
         $^{24}$Na & $1^+_1\rightarrow 0^+_1$ &5.80 &5.188--5.223 &4.448--4.545\\
         $^{24}$Na & $4^+_{\rm gs}\rightarrow 4^+_1$ &6.11  &5.416--5.461 &5.795--5.866\\ 
         $^{24}$Na & $4^+_{\rm gs}\rightarrow 3^+_1$ &6.60  &5.727--5.773 &6.342--6.422\\ 
    \end{tabular}
    \end{ruledtabular}
        \label{tab:beta_decays}
    \end{threeparttable}
\end{table}

We also compare the VS-IMSRG results with those of phenomenological NSM calculations performed using the NuShellX@MSU code \cite{Brown2014} in the $sd$-shell with the USDB interaction \cite{Brown2006}.
This interactions is based on a renormalized $G$-matrix with two-body matrix elements adjusted to experimental binding and excitation energies of $sd$-shell. Hence, the interaction generally well reproduces the spectroscopic properties of $sd$-shell nuclei.
As NSM is a phenomenological method, the fitted Hamiltonian and bare OMC operators are inconsistent. 
For the NSM calculations, we use a HO basis with frequency obtained from the Blomqvist-Molinari formula \cite{Blomqvist1968} $\hbar\omega=(45A^{-1/3}-25A^{-2/3})~{\rm MeV}$ for evaluating the operator matrix elements. 
However, the single-particle basis is based on the USDB interaction and thereby differs from the HO basis. 
Furthermore, in the NSM we are restricted to the $sd$-shell, and contributions to OMC from outside the valence space are not accounted for with the bare transition operator.

\section{\label{sec:results}Results and Discussion}

\subsection{Spectroscopy of $^{24}$Mg and $^{24}$Na}

In order to test the validity of the chosen many-body methods, we compare calculated nuclear observables with experimental data, where available. In Fig. \ref{fig:excitation_energies}, we plot the computed excitation energies in the nuclei of interest against the experimental spectra. We only show the positive parity states - that is, the states that can be obtained within the $sd$ shell. While excitation energies have a negligible direct impact on the capture rate through \eqref{eq:q}, the energies are often used as an indirect probe of the quality of the shell-model calculations.
In Fig. \ref{fig:excitation_energies}, we see that the NSM better reproduces the experimental excitation energies, particularly in the odd-odd system. 
This is expected, since the USDB interaction is tuned to reproduce spectroscopic properties of $sd$-shell nuclei, whereas the 1.8/2.0 (EM) chiral Hamiltonian used in the VS-IMSRG is only informed by few-body data.
Consequently, the possibility of overconstraining excitation energies (at the expense of observables like OMC, which are not included in the fit) means this proxy should not be given undue weight.

In Table \ref{tab:electromagnetic_moments}, we compare calculated magnetic dipole and electric quadrupole moments with existing experimental data. For NSM, we use the effective charges $e_{\rm eff}^{\rm p}=1.5$ and $e_{\rm eff}^{\rm n}=0.5$ and the $g$ factors $g_l^{\rm p}=1.0$, $g_{l}^{\rm n}=0.0$, $g_{s}^{\rm eff}({\rm p})=3.910$ and $g_s^{\rm eff}({\rm n})=-2.678$, while for VS-IMSRG, we use bare values. 
We notice that for the magnetic dipole moments ($\mu$), VS-IMSRG generally does well in $^{24}$Mg, whereas for $^{24}$Na NSM is clearly in better agreement with experiment. 
In particular, for the $1^+_{\rm g.s.}$ VS-IMSRG agrees poorly with the experimental counterpart.  
For electric quadrupole moments ($Q$), there is little experimental data to compare, but we note both methods underestimate the absolute value of $Q(2^+_1)$ in $^{24}$Mg -- VS-IMSRG more notably. 
We also compare the computed $B(E2)$ and $B(M1)$ values with experiment in Table \ref{tab:magnetic_decays}. 
As can be seen in the table, VS-IMSRG describes the $M1$ transitions slightly better than NSM, but underestimates the $E2$ transition strengths.
This is a well-known consequence of the VS-IMSRG(2) approximation \cite{Hend18E2,Stroberg2022}, and we would expect improvement with IMSRG(3).

Lastly, in Table~\ref{tab:beta_decays} we compare calculated $\log ft$ values for $\beta$ decays of $^{24}$Na leading to excited states in $^{24}$Mg. For NSM, we scale the involved Gamow-Teller NMEs by the typical zero-momentum-exchange quenching %$0.820\pm0.015$ for the $p$-shell nuclei \cite{Chou1993} and
$0.77\pm 0.02$ \cite{Wildenthal1983}, while for the VS-IMSRG we include the two-body currents by correcting $g_{\rm A}(q^2)$ by $\delta_a(0)$ introduced in Section \ref{ssec:twobodycurrents}. 
For VS-IMSRG, the ranges shown in the table also contain the uncertainty coming from the choice of reference state in the calculations. The reference-state dependence is much smaller than the uncertainty of the two-body currents.
Generally, VS-IMSRG describes the measured $\log ft$ values better than NSM. However, VS-IMSRG notably underestimates the $\log ft$ of the transition $^{24}{\rm Na}(1^+_1)\rightarrow~^{24}{\rm Mg}(0^+_1)$, likely due to mixing of $1^+$ states.

\subsection{Nuclear Matrix Elements for Muon Capture}
\begin{figure*}[h!]
    \centering
    \includegraphics[width=\linewidth]{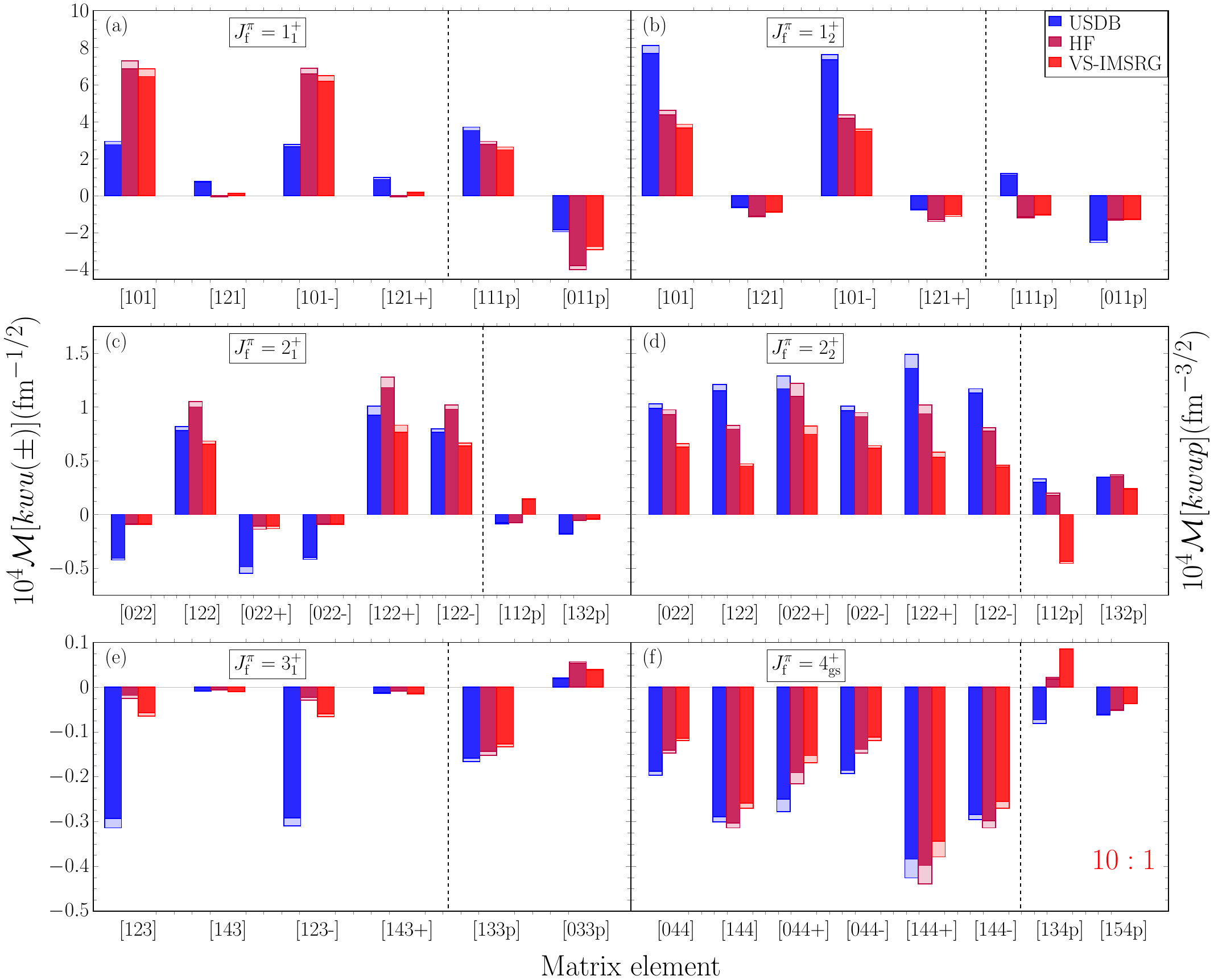}
    \caption{Nuclear matrix elements for the transitions $\mu^-+~^{24}{\rm Mg}(0^+_{\rm g.s.})\rightarrow \nu_{\mu}+~^{24}{\rm Na}(J^{\pi}_i)$ to a few lowest $J^{\pi}$ states in $^{24}$Na. The dark bars are computed with the Dirac wave functions and the light bars with the BS approximation. Thus, the difference shows the finite-size effect. The matrix elements of the type $\mathcal{M}[kwup]$ -- separated by the vertical dashed line -- are given in $\rm fm ^{-3/2}$, while the rest are in units $\rm fm ^{-1/2}$. The NMEs in panel (f) are multiplied by 10.}
    \label{fig:24Na-all-NMEs}
\end{figure*}

\begin{figure*}[h!]
    \centering
    \includegraphics[width=\linewidth]{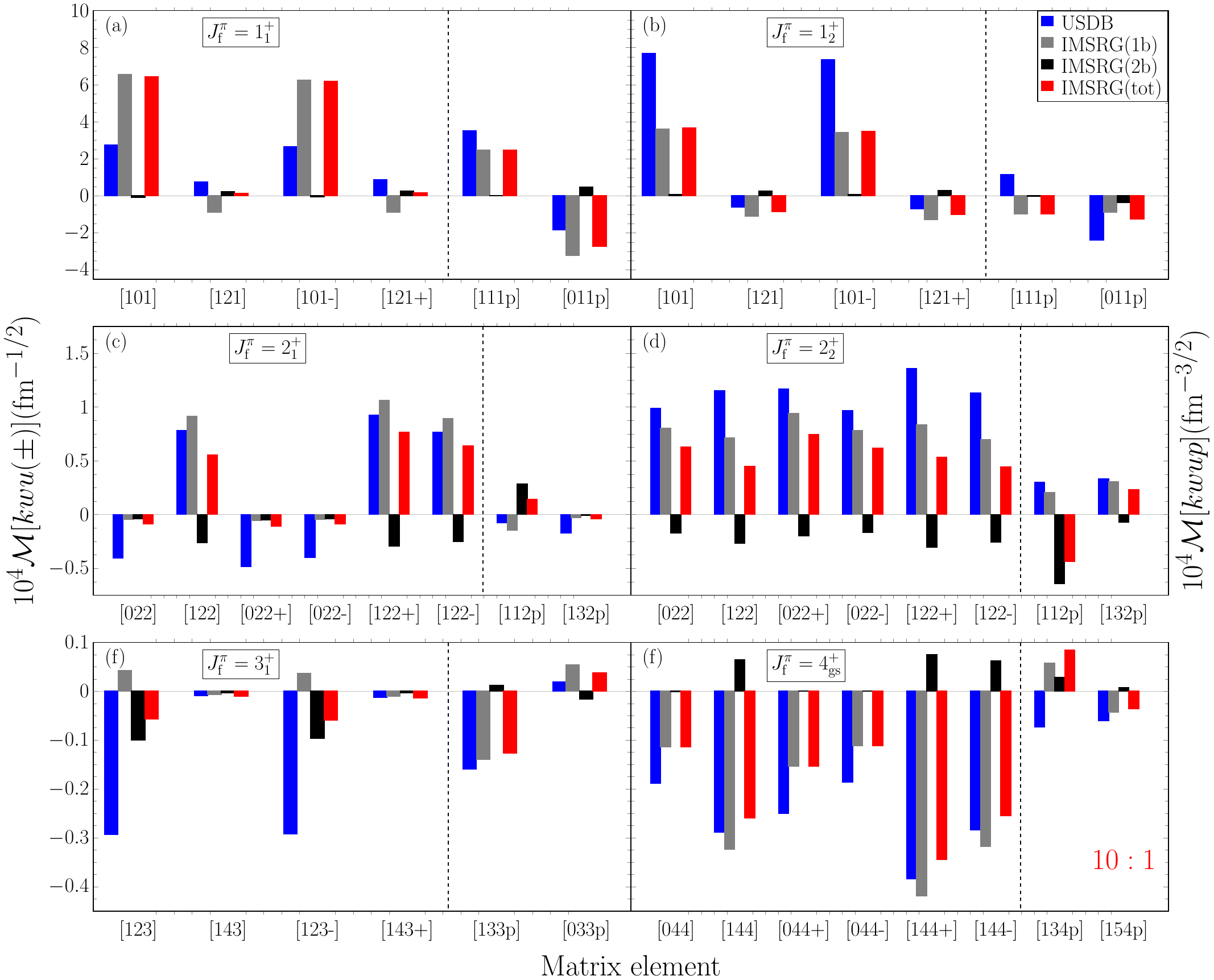}
    \caption{One-body (gray bars) and two-body (black bars) components of the total VS-IMSRG NMEs (red bars) for $\mu^-+~^{24}{\rm Mg}(0^+_{\rm g.s.})\rightarrow \nu_{\mu}+~^{24}{\rm Na}(J^{\pi}_i)$ compared with USDB NMEs (blue bars). NMEs are computed with the exact bound-muon wave function with finite-size nucleus. The NMEs in panel (f) are multiplied by 10.}
    \label{fig:24Na-all-one-two}
\end{figure*}

In Fig.~\ref{fig:24Na-all-NMEs}, we show the computed OMC NMEs for the transitions to the lowest states in $^{24}$Na. 
We compare the NMEs computed in the VS-IMSRG framework with those obtained in the NSM with the USDB interaction and OMC operators expressed in the HO basis. 
The $q$-values in the operators correspond to the experimental excitation energies. 
The `HF' in the figures refers to the matrix elements calculated with the decoupled VS-IMSRG Hamiltonian and the operators evaluated in the Hartree-Fock basis (i.e. without consistent IMSRG evolution). 
Hence, the HF results do not contain the two-body terms induced in the VS-IMSRG procedure. 
The dark bars for each framework are computed with the bound-muon wave function solved from the Dirac equation taking into account the finite size of the nucleus, whereas the light bars are computed with the BS pointlike-nucleus approximation introduced in Sec.~\ref{ssec:muonwavefcns}. 
Hence, the difference shows the effect coming from finite-size nucleus. 
Note that the momentum-dependent matrix elements of the type $\mathcal{M}[kwup]$ are given in units $\rm fm ^{-3/2}$ here, while the rest are given in units $\rm fm ^{-1/2}$.

The finite-size effect is rather consistent for all NMEs: taking this into account decreases the absolute value by $5-10\%$. 
The finite-size effect is slightly larger in the NMEs of the type $[kwu+]$, whereas the effect is somewhat less visible for $[kwu-]$. 
These small differences can be explained by the derivative terms in $[kwu\pm]$ (see Table \ref{tab:operators}) that are not present in the other matrix elements. 
Note that the finite-size effect on the matrix elements is similar to that on the wave function at $r=0$ in Fig.~\ref{fig:muonwavefunctions}.

The `HF' bars can be considered an intermediate step between the `USDB' and `VS-IMSRG' results.
In most cases the `HF' NMEs are closer to `VS-IMSRG' than `USDB'. 
This means that in most cases, transforming the OMC operators and including the two-body terms have a relatively small effect on the NMEs. 
However, particularly for transitions to $2_{1,2}^+$ states, including the two-body term has a strong effect on the $[112p]$ NME: including the two-body term in `VS-IMSRG' changes the sign (see panels (c) and (d) in Fig.~\ref{fig:24Na-all-NMEs}). 
In order to further study the effects coming from the IMSRG transformation and the induced two-body part, we  compare the `USDB' NMEs against the one- and two-body parts and the total IMSRG matrix elements in Fig.~\ref{fig:24Na-all-one-two}. 
These figures confirm the perceptions mentioned above: the IMSRG two-body term indeed is sizeable but of the opposite sign especially in the case of the $[112p]$ NME, as can be seen in panels (c) and (d) of Fig.~\ref{fig:24Na-all-one-two}. 
In addition, for the transition to $4^+_{\rm g.s.}$, shown in panel (f) of Fig.~\ref{fig:24Na-all-one-two}, we see that the two-body term increases the value of the $[134p]$ NME by $\approx 50\%$.

On the other hand, we note that using the VS-IMSRG wave functions instead of those of the NSM has a strong effect on the NMEs. 
This can be seen if we compare the `USDB' bars with the `HF' or `VS-IMSRG'. This stems from the fact that in the VS-IMSRG calculations, the operators and excitations are derived consistently with the same unitary transformation, while in the NSM calculations, the excitations outside the valence space are implicitly accounted for by the effective interaction but the operator is not adjusted correspondingly. 
In order to distinguish effects coming from the different wave functions versus transforming the operator, we show the one-body parts of the NMEs obtained with different OBTD/operator combinations: we use either the USDB or the IMSRG OBTDs together with either the HO operator or the IMSRG-transformed operator. 
The results are given in Tables \ref{tab:1+_1}-\ref{tab:4+_gs} in the Appendix. 
In most cases, the transformed operator does not significantly change the NME, while the OBTD has a more pronounced effect: in many cases the NMEs obtained with different OBTDs are opposite in sign.
Interestingly, the effect of the VS-IMSRG on the matrix element $[101]$ (and $[101-]$, which differs from $[101]$ only by the small derivative term) is the opposite for the states $1^+_1$ and $1^+_2$: where the VS-IMSRG result is larger than the USDB one for the first $1^+$ state, the situation is the opposite for the second $1^+$ state, likely due to mixing of the two states.

\begin{figure*}[h!]
\begin{subfigure}[t]{0.45\linewidth}
    \centering
    \includegraphics[width=\linewidth]{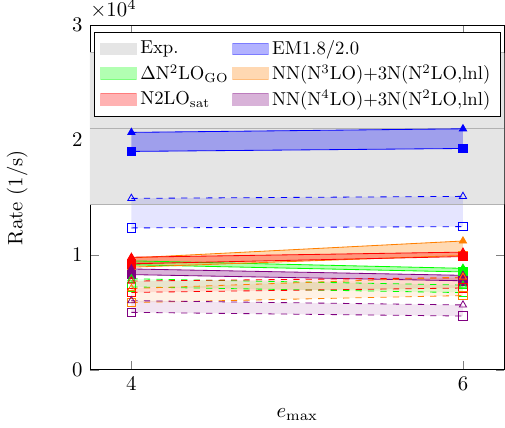}
    \caption{$J^{\pi}_f=1^+_1$}
\end{subfigure}
\begin{subfigure}[t]{0.45\linewidth}
    \centering
    \includegraphics[width=\linewidth]{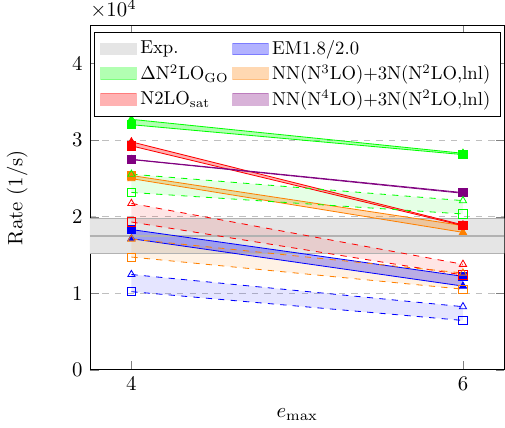}
    \caption{$J^{\pi}_f=1^+_2$}
\end{subfigure}
\begin{subfigure}[t]{0.45\linewidth}
    \centering
    \includegraphics[width=\linewidth]{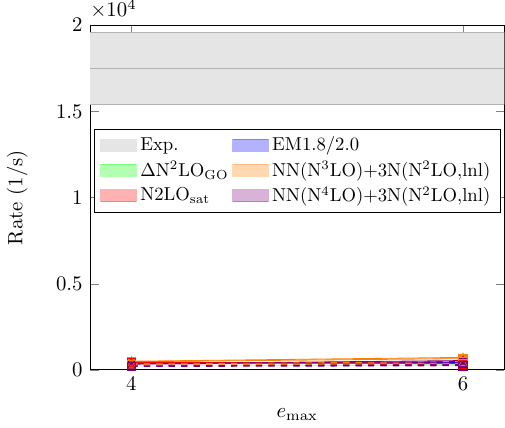}
    \caption{$J^{\pi}_f=2^+_1$}
\end{subfigure}
\begin{subfigure}[t]{0.45\linewidth}
    \centering
    \includegraphics[width=\linewidth]{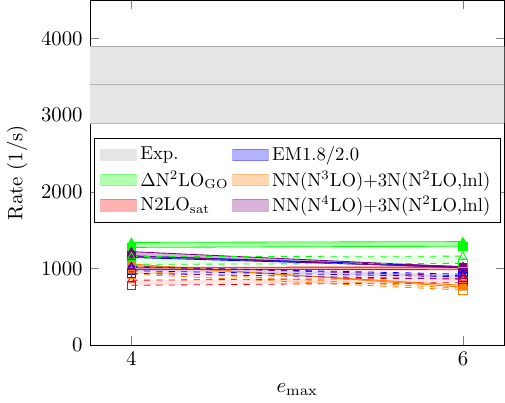}
    \caption{$J^{\pi}_f=2^+_2$}
    \end{subfigure}
    \caption{Capture rates to the different $J^{\pi}_f$ states in $^{24}$Na obtained with different interactions compared with the experimental value. The lighter bands include the effect of 2BCs.}
    \label{fig:24Na_diff_int}
\end{figure*}

The reference-state dependence in the case of muon-capture NMEs is $\lesssim 2\%$. The results presented in the tables and figures are obtained with the $^{24}$Na  reference state. Here we only show the results for the interaction EM1.8/2.0. In order to study the interaction dependence of the VS-IMSRG computed muon-capture rates, we compare the rates obtained with different chiral interactions: $\Delta$N$^{2}$LO$_{\rm GO}$(394) \cite{Jiang2020}, N2LO$_{\rm sat}$ \cite{Ekstrom2015}, NN(N$^{3}$LO)~\cite{Entem2003} + 3N(N$^{2}$LO, lnl)~\cite{Soma2020}, and NN(N$^{4}$LO)~\cite{Entem2017} + 3N(N$^{2}$LO, lnl)~\cite{Gysbers2019}, against the experimental rates in Fig. \ref{fig:24Na_diff_int}. 
It can be seen that none of the employed interactions is sufficient to reproduce the measured capture rates to all the studied excited states in $^{24}$Na: where EM1.8/2.0 does a better job for the first $1^+$ state, it underestimates the rate to the second $1^+$ state while the other interactions give better estimates for it. On the other hand, none of the interactions is capable of describing the capture rates to the $2^+$ excited states. We find that the interaction dependence ($\sim 5\%$) is larger than many-body calculation uncertainty, and therefore the interaction sensitivity needs to be assessed for an improved understanding.

\subsection{Capture Rates}
In Table~\ref{tab:rates-exp-comparison}, we give the capture rates to the lowest states in $^{24}$Na obtained from the calculated NMEs. For each state we give the capture rates obtained from the NSM and VS-IMSRG calculations with and without two-body currents, compared against experimental data.  
The capture rates obtained with the BS pointlike-nucleus approximation are shown in parentheses, while the rest of the rates are obtained with the realistic bound-muon wave functions. 

Comparing the obtained capture rates in columns 4-7 of Table~\ref{tab:rates-exp-comparison} we see that the OMC rates for the VS-IMSRG and the NSM show an overall consistent pattern for all states except the $1^+$ states which seem to be interchanged between the two calculations. 
The only notable difference in the magnitudes of the OMC rates, excluding the $1^+$ states, concerns the OMC to the $3^+_1$ state, with a factor of 20 difference. Concerning the $1^+$ states, the behavior of the matrix elements $[101]$ and $[101-]$ is reflected in the capture rates: VS-IMSRG predicts a notably larger rate to $1^+_1$ and a smaller rate to $1^+_2$ compared to the NSM. 
This is reasonable since the $[101]$ and $[101-]$ NMEs are dominant for these transitions. 

The finite-size effect of the NMEs is shown to be rather constant, reducing the capture rates by $10-20\%$, in keeping with the finite-size effect of the NMEs.
The effect is smaller than the scaling factor $(Z_{\rm eff}/Z)^4=(10.69/12)^4\approx 0.63$ (for $^{24}$Mg) \cite{Suzuki1987} used in the previous studies to account for the finite-size effect. Note that the bound-muon wave function of the present study corresponds to $Z_{\rm eff}=10.96$, which would give a larger scaling factor of 0.70. The difference in the effective atomic numbers is partly explained by different charge radii: $Z_{\rm eff}$ of \cite{Suzuki1987} assumes $R_c=1.69A^{1/3}$ fm instead of $R_c=1.2A^{1/3}$ fm.
The hadronic two-body currents, on the other hand, reduce the capture rates by $20-30\%$. The reduction is mainly due to the axial-vector two-body current, but partially balanced by the non-zero effect of the pseudoscalar two-body current at finite-momentum transfer.

Finally, we compare the computed capture rates to the $1^+$ and $2^+$ states with the partial muon capture rates reported in Ref.~\cite{Gorringe1999}. 
They measured the direct (produced by muon-capture on the ground state of $^{24}$Mg) and indirect (produced by muon-capture on an excited state of $^{24}$Mg) $\gamma$-ray yields to a few low-lying states in $^{24}$Na. 
The partial muon-capture rates can be obtained by multiplying the direct state yields by the appropriate total muon-capture rates~\cite{Suzuki1987}. 
The obtained values, together with the capture rates obtained in the present work, are tabulated in Table \ref{tab:rates-exp-comparison}. It should be noted, however, that obtaining the direct-state yields requires knowledge of cascade feeding, and only a small fraction of the $\gamma$-rays could be identified in Ref.~\cite{Gorringe1999}.
Therefore all values could be slightly reduced following the discovery of more transitions. 
As for the $1^+_1$ state, the authors suspected that the cascade feeding was unidentified.

\begin{table*}
    \caption{Comparison between the computed capture rates with (1BC+2BC) or without (1BC) the effect of two-body currents and the experimental partial muon capture rates obtained from Ref.~\cite{Gorringe1999} (see text) to the low-lying $1^+$ and $2^+$ states in $^{24}$Na. The rates computed with point-like-nucleus approximation are shown in parenthesis. The uncertainties of the summed rates are obtained by summing up uncertainties of individual capture rates.}
    \label{tab:rates-exp-comparison}
    \begin{threeparttable}
\begin{ruledtabular}
    \centering
    \begin{tabular}{cccccccc}
    $J^{\pi}_i$ &$E_{\rm exp}$ (MeV)&\multicolumn{5}{c}{Rate ($10^3$ 1/s)}\\
    \cline{3-7}
     & & Exp. & \multicolumn{2}{c}{NSM} & \multicolumn{2}{c}{IMSRG}\\
     \cline{4-5}\cline{6-7}
     & &\cite{Gorringe1999} &1BC &1BC+2BC &1BC &1BC+2BC\\
     \hline
         $1^+_1$ & 0.472 &$21.0\pm 6.6$\tnote{a} &4.0 (4.5)  &2.9--3.1 (3.2--3.5) &22.3 (25.2) & 14.3--16.0 (16.1--18.0)\\ 
         $1^+_2$ & 1.347 &$17.5\pm 2.3$ &32.7 (36.3) &20.2--22.8 (22.5--25.3) &7.7 (8.5) &4.5--5.2 (5.0--5.7)\\
         Sum($1^+$) &  &$38.5\pm 8.9$ &36.7 (40.8) &23.1--25.9 (25.7--28.8) &30.0 (33.7) &18.8--21.2 (21.1--23.7)\\ 
         $2^+_1$ & 0.563  &$17.5\pm 2.1$  &1.0 (1.0) &0.7--0.7 (0.7--0.8) &0.5 (0.5) & 0.3--0.3 (0.3--0.4)\\ 
         $2^+_2$ & 1.341  &$3.4\pm 0.5$ &3.1 (3.4) &2.4--2.6 (2.7--2.8)  &1.0 (1.1)&0.9--0.9 (1.0--1.0)\\ 
         Sum($2^+$) &   &$20.9\pm 2.6$ &4.1 (4.4) &3.1--3.3 (3.4--3.6) &1.5 (1.6) &1.2--1.2 (1.3--1.4)\\
         $4^+_{\rm gs}$ &0.0 & - & 0.003 (0.003) &0.002--0.002 (0.002--0.003) &0.002 (0.002) & 0.001--0.002 (0.002--0.002)\\ 
         $3^+_1$ &1.345 &- &0.11 (0.13) &0.07--0.08 (0.08--0.09) &0.005 (0.006) &0.004--0.004 (0.005--0.005)\\ 
    \end{tabular}
    \end{ruledtabular}
    \begin{tablenotes}
      \item[a]{Unidentified cascade feeding of the state is suspected.}
    \end{tablenotes}
    \end{threeparttable}
\end{table*}

We note that while the capture rates to the lowest two $1^+$ states in Table \ref{tab:rates-exp-comparison} seem to be interchanged in the NSM and VS-IMSRG, the summed rates to these two states agree within 20\%. 
Clearly both of the estimates are notably smaller than the experimental counterpart, which could be explained by potentially unidentified $\gamma$-rays. 
For the $2^+$ states the differences are more prominent: the NSM- and VS-IMSRG-calculated summed rates disagree almost by a factor of three, while the experimental value is an order of magnitude larger than the computed values. 
Most of the difference appears to be coming from the capture rates to the $2^+_1$ state, where the theory predictions are $\approx 20-40$ times smaller than the measured rate. 
The reason for these discrepancies is largely explained by the strong interaction dependence of the rates, and will likely be illuminated by upcoming experiments.

\section{\label{sec:Summary}Summary and Outlook}

In this work, we study OMC on $^{24}$Mg generating transitions to the lowest excited states in $^{24}$Na using the NSM and VS-IMSRG. The study of the spectroscopic properties of both nuclei reveals that most of the available spectroscopic data are reasonably well described by both nuclear models. Encouraged by this, we proceed to compute the OMC rates by calculating the OMC NMEs based on Morita-Fujii
muon-capture formalism and apply the VS-IMSRG ab initio framework to obtain consistent valence-space Hamiltonians and OMC operators. 
In addition, we use realistic bound-muon wave functions obtained from solving the Dirac equations where the finite size of the nucleus is taken into account. 
Finally, we calculate the capture rates to the lowest states in $^{24}$Na with the obtained matrix elements and compare them against existing experimental data.

Comparing the VS-IMSRG results with those of the NSM, we see that explicitly including excitations outside the valence space in the form of the OBTDs generally has strong impact on the NMEs, while the effect coming from consistent transformation of the OMC operators is less significant, with a few exceptions.
While we anticipate including IMSRG(3) level corrections to the operators would have a minor influence on the NMEs, such corrections to the effective valence-space Hamiltonian could be important.
The VS-IMSRG-computed capture rates are generally smaller than the rates computed in NSM, but the rates to the first and second $1^+$ states seem to be interchanged, which is partly explained by the mixing of these two states.
Comparing with experimental data, we notice that the obtained capture rates are generally smaller than the experimental rates; while the agreement is reasonable for the total transition rate to the $1^+$ states, both the NSM and VS-IMSRG underestimate the total rate to the $2^+$ states. 
The discrepancy could be partly explained by uncertainties in the experimental data, but future measurements will help shed light on this.

The present work serves as the first step towards a systematic ab initio treatment of OMC on nuclei. The present results, compared with the available OMC data leave room for improvement and it remains as a future task to study further theoretical refinements such as including
the small component of the bound-muon wave function and some higher-order corrections to IMSRG, which would hopefully improve the description of muon capture on nuclei. 
We are currently extending these studies to all nuclei relevant for upcoming experiments in order to investigate whether similar difficulties in predicting the OMC rates in both nuclear models appear and in order to test further refinements of the OMC calculations. The ultimate goal is to
compare the obtained partial capture rates with experimental data to shed light on effective values of the couplings $g_{\rm A}$ and $g_{\rm P}$ at momentum exchange $q\sim 100~{\rm MeV}$. 
This momentum regime is highly relevant for $0\nu\beta\beta$ decay and thus OMC calculations may help constrain uncertainties related to emerging ab initio predictions of $0\nu\beta\beta$ decay~\cite{Yao20,Bell21Ge,Novario20}.

\clearpage
\begin{acknowledgments}

We thank J. Men\'{e}ndez and B. S. Hu for helpful discussions.
This work was supported by the Finnish Cultural Foundation grant No. 00210067, NSERC under grants SAPIN-2018-00027 and RGPAS-2018-522453, the Arthur B. McDonald Canadian Astroparticle Physics Research Institute, the US Department of Energy (DOE) under contracts DE-FG02-97ER41014 and DE-AC02-06CH11357, Academy of Finland (Grant Nos. 314733, 320062, 345869 and 318043), and the Deutsche Forschungsgemeinschaft (DFG, German Research Foundation) -- Project-ID 279384907 -- SFB 1245.

TRIUMF receives funding via a contribution through the National Research Council of Canada.
Computations were performed with an allocation of computing resources on Cedar at WestGrid and Compute Canada, and on the Oak Cluster at TRIUMF managed by the University of British Columbia department of Advanced Research Computing (ARC).

\end{acknowledgments}

\bibliography{24Mg-bib}

\appendix
\section{}
In Table \ref{tab:2BCs}, we list the values of the two-body corrections obtained using the LECs corresponding to the EM1.8/2.0 interaction for each of the transitions considered in the present work.
\begin{table}[h!]
    \centering
    \caption{The values of $\delta_a(q^2)$ and $\delta_a^P(q^2)$ obtained with the LECs listed in Table \ref{tab:LECs} for the NSM and IMSRG calculations.}
    \begin{ruledtabular}
    \begin{tabular}{cccccc}
          Final state &$q$(MeV) &$\delta_a(q^2)$ &$\delta_a^P(q^2)$\\
         \hline
         $^{24}{\rm Na}(4^+_{\rm gs})$ &99.357 &$-0.271\dots -0.211$ &$0.144\dots 0.178$\\
         $^{24}{\rm Na}(1^+_1)$ &98.886 &$-0.270\dots -0.211$ &$0.143\dots 0.177$\\
         $^{24}{\rm Na}(2^+_1)$ &98.759 &$-0.270\dots -0.211$ &$0.143\dots 0.177$\\
         $^{24}{\rm Na}(2^+_2)$ &98.019 &$-0.270\dots -0.211$ &$0.142\dots 0.175$\\
         $^{24}{\rm Na}(3^+_1)$ &98.015 &$-0.270\dots -0.211$ &$0.142\dots 0.175$\\
         $^{24}{\rm Na}(1^+_2)$ &98.013 &$-0.270\dots -0.211$ &$0.142\dots 0.175$\\
    \end{tabular}
    \end{ruledtabular}
    \label{tab:2BCs}
\end{table}

To evaluate the relative importance of the different (NSM/VS-IMSRG) nuclear wave functions and operators (HO/VS-IMSRG -evolved), we show the NMEs obtained with different (OBTD, operator)-combinations in Tables~\ref{tab:1+_1}-\ref{tab:4+_gs}. 
The HO operators are obtained with $\hbar\omega$=$16~\rm MeV$, the same value that was used in the VS-IMSRG evolution. 
Hence, the (USDB,HO) combinations do not exactly correspond to the `USDB' NMEs shown in Fig.~\ref{fig:24Na-all-NMEs} and \ref{fig:24Na-all-one-two}, where the Blomqwist-Molinari formula was used. 
The (IMSRG,IMSRG) combinations in the tables correspond to the VS-IMSRG(1b) NMEs shown in Fig.~\ref{fig:24Na-all-one-two}.

\begin{table*}[h!]
\centering
 \caption{The NMEs for the transition $\mu^-+~^{24}{\rm Mg}(0^+_{\rm g.s.})\rightarrow \nu_{\mu}+~^{24}{\rm Na}(1^+_1)$ obtained with different (OBTD, operator) -combinations.}
    \begin{ruledtabular}
    \begin{tabular}{cccccccc}
OBTD	&operator	&[101]	    &[121]	    &[101-]	    &[121+]	    &[111p]	    &[011p]   \\
\hline
USDB	&HO	       &3.11$\times10^{-4}$	&6.32$\times10^{-5}$	&3.01$\times10^{-4}$	&7.58$\times10^{-5}$	&3.79$\times10^{-4}$	&-1.98$\times10^{-4}$\\
USDB	&IMSRG	   &3.32$\times10^{-4}$	&6.34$\times10^{-5}$	&3.21$\times10^{-4}$	&7.54$\times10^{-5}$	&3.18$\times10^{-4}$	&-1.82$\times10^{-4}$\\
IMSRG	&HO		   &7.68$\times10^{-4}$	&4.51$\times10^{-6}$	&7.41$\times10^{-4}$	&5.75$\times10^{-6}$	&2.88$\times10^{-4}$	&-3.82$\times10^{-4}$\\
IMSRG	&IMSRG	   &6.54$\times10^{-4}$	&-8.92$\times10^{-6}$	&6.25$\times10^{-4}$	&-8.96$\times10^{-6}$	&2.46$\times10^{-4}$	&-3.22$\times10^{-4}$\\
    \end{tabular}
    \end{ruledtabular}
    \label{tab:1+_1}
\end{table*}

\begin{table*}[h!]
\centering
 \caption{The same as in Table \ref{tab:1+_1} but for transition $\mu^-+~^{24}{\rm Mg}(0^+_{\rm g.s.})\rightarrow \nu_{\mu}+~^{24}{\rm Na}(1^+_2)$.}
    \begin{ruledtabular}
    \begin{tabular}{cccccccc}
OBTD	&operator	&[101]	    &[121]	    &[101-]	    &[121+]	    &[111p]	    &[011p]   \\
\hline
USDB	&HO		&8.69$\times10^{-4}$	&-5.18$\times10^{-5}$	&8.38$\times10^{-4}$	&-6.16$\times10^{-5}$	&1.22$\times10^{-4}$	&-2.61$\times10^{-4}$\\
USDB	&IMSRG	&7.08$\times10^{-4}$	&-7.96$\times10^{-5}$	&6.76$\times10^{-4}$	&-9.18$\times10^{-5}$	&1.12$\times10^{-4}$	&-1.94$\times10^{-4}$\\
IMSRG	&HO		&4.98$\times10^{-4}$	&-8.63$\times10^{-5}$	&4.79$\times10^{-4}$	&-1.03$\times10^{-4}$	&-1.22$\times10^{-4}$	&-1.40$\times10^{-4}$\\
IMSRG	&IMSRG	&3.61$\times10^{-4}$	&-1.11$\times10^{-4}$	&3.42$\times10^{-4}$	&-1.29$\times10^{-4}$	&-9.63$\times10^{-5}$	&-8.72$\times10^{-5}$\\
    \end{tabular}
    \end{ruledtabular}
    \label{tab:1+_2}
\end{table*}

\begin{table*}[h!]
\centering
 \caption{The same as in Table \ref{tab:1+_1} but for transition $\mu^-+~^{24}{\rm Mg}(0^+_{\rm g.s.})\rightarrow \nu_{\mu}+~^{24}{\rm Na}(2^+_1)$.}
    \begin{ruledtabular}
    \begin{tabular}{cccccccccc}
OBTD	&operator	&[022]	    &[122]	    &[022+]	    &[022-]	    &[122+]	    &[122-]	    &[112p]	    &[132p]   \\
\hline
USDB	&HO		&-3.41$\times10^{-5}$	&6.65$\times10^{-5}$	&-4.11$\times10^{-5}$	&-3.36$\times10^{-5}$	&7.95$\times10^{-5}$	&6.53$\times10^{-5}$	&-6.46$\times10^{-6}$	&-1.45$\times10^{-5}$\\
USDB	&IMSRG	&-3.11$\times10^{-5}$	&7.42$\times10^{-5}$	&-3.69$\times10^{-5}$	&-3.04$\times10^{-5}$	&8.62$\times10^{-5}$	&7.23$\times10^{-5}$	&-2.12$\times10^{-5}$	&-1.27$\times10^{-5}$\\
IMSRG	&HO		&-5.35$\times10^{-6}$	&7.95$\times10^{-5}$	&-6.83$\times10^{-6}$	&-5.36$\times10^{-6}$	&9.48$\times10^{-5}$	&7.80$\times10^{-5}$	&1.84$\times10^{-6}$	&-4.27$\times10^{-6}$\\
IMSRG	&IMSRG	&-4.69$\times10^{-6}$	&9.17$\times10^{-5}$	&-5.81$\times10^{-6}$	&-4.65$\times10^{-6}$	&1.06$\times10^{-4}$	&8.92$\times10^{-5}$	&-1.46$\times10^{-5}$	&-3.16$\times10^{-6}$\\
    \end{tabular}
    \end{ruledtabular}
    \label{tab:2+_1}
\end{table*}

\begin{table*}[h!]
\centering
 \caption{The same as in Table \ref{tab:1+_1} but for transition $\mu^-+~^{24}{\rm Mg}(0^+_{\rm g.s.})\rightarrow \nu_{\mu}+~^{24}{\rm Na}(2^+_2)$.}
    \begin{ruledtabular}
    \begin{tabular}{cccccccccc}
OBTD	&operator	&[022]	    &[122]	    &[022+]	    &[022-]	    &[122+]	    &[122-]	    &[112p]	    &[132p]   \\
\hline
USDB	&HO		&8.35$\times10^{-5}$	&9.82$\times10^{-5}$	&1.00$\times10^{-4}$	&8.20$\times10^{-5}$	&1.17$\times10^{-4}$	&9.64$\times10^{-5}$	&2.58$\times10^{-5}$	&2.83$\times10^{-5}$\\
USDB	&IMSRG	&7.55$\times10^{-5}$	&1.08$\times10^{-4}$	&8.93$\times10^{-5}$	&7.38$\times10^{-5}$	&1.26$\times10^{-4}$	&1.05$\times10^{-4}$	&1.32$\times10^{-5}$	&2.70$\times10^{-5}$\\
IMSRG	&HO		&8.31$\times10^{-5}$	&6.82$\times10^{-5}$	&9.97$\times10^{-5}$	&8.16$\times10^{-5}$	&8.16$\times10^{-5}$	&6.70$\times10^{-5}$	&2.26$\times10^{-5}$	&3.03$\times10^{-5}$\\
IMSRG	&IMSRG	&8.01$\times10^{-5}$	&7.16$\times10^{-5}$	&9.43$\times10^{-5}$	&7.82$\times10^{-5}$	&8.37$\times10^{-5}$	&6.98$\times10^{-5}$	&2.07$\times10^{-5}$	&3.04$\times10^{-5}$\\
    \end{tabular}
    \end{ruledtabular}
    \label{tab:2+_2}
\end{table*}

\begin{table*}[h!]
\centering
 \caption{The same as in Table \ref{tab:1+_1} but for transition $\mu^-+~^{24}{\rm Mg}(0^+_{\rm g.s.})\rightarrow \nu_{\mu}+~^{24}{\rm Na}(3^+_1)$.}
    \begin{ruledtabular}
    \begin{tabular}{cccccccc}
OBTD	&operator	&[123]	    &[143]	    &[123-]	    &[143+]	    &[133p]	    &[033p]   \\
\hline
USDB	&HO		&-2.32$\times10^{-5}$	&-5.99$\times10^{-7}$	&-2.32$\times10^{-5}$	&-8.17$\times10^{-7}$	&-1.32$\times10^{-5}$	&1.47$\times10^{-6}$\\
USDB	&IMSRG	&-4.93$\times10^{-6}$	&-5.24$\times10^{-7}$	&-4.65$\times10^{-6}$	&-7.46$\times10^{-7}$	&-1.24$\times10^{-5}$	&-1.60$\times10^{-7}$\\
IMSRG	&HO		&-7.61$\times10^{-6}$	&-7.59$\times10^{-7}$	&-7.69$\times10^{-6}$	&-1.03$\times10^{-6}$	&-1.30$\times10^{-5}$	&4.80$\times10^{-6}$\\
IMSRG	&IMSRG	&4.27$\times10^{-6}$	&-6.28$\times10^{-7}$	&3.48$\times10^{-6}$	&-8.70$\times10^{-7}$	&-1.24$\times10^{-5}$	&4.80$\times10^{-6}$\\
    \end{tabular}
    \end{ruledtabular}
    \label{tab:3+_1}
\end{table*}

\begin{table*}[h!] 
\centering
 \caption{The same as in Table \ref{tab:1+_1} but for transition $\mu^-+~^{24}{\rm Mg}(0^+_{\rm g.s.})\rightarrow \nu_{\mu}+~^{24}{\rm Na}(4^+_{\rm g.s.})$.}
    \begin{ruledtabular}
    \begin{tabular}{cccccccccc}
OBTD	&operator	&[044]	    &[144]	    &[044+]	    &[044-]	    &[144+]	    &[144-]	    &[134p]	    &[154p]   \\
\hline
USDB	&HO		&-1.24$\times10^{-6}$	&-1.90$\times10^{-6}$	&-1.68$\times10^{-6}$	&-1.23$\times10^{-6}$	&-2.57$\times10^{-6}$	&-1.88$\times10^{-6}$	&-4.89$\times10^{-7}$	&-3.99$\times10^{-7}$\\
USDB	&IMSRG	&-8.93$\times10^{-7}$	&-3.14$\times10^{-6}$	&-1.22$\times10^{-6}$	&-8.83$\times10^{-7}$	&-4.07$\times10^{-6}$	&-3.08$\times10^{-6}$	&6.48$\times10^{-7}$	&-3.54$\times10^{-7}$\\
IMSRG	&HO		&-1.38$\times10^{-6}$	&-1.96$\times10^{-6}$	&-1.87$\times10^{-6}$	&-1.37$\times10^{-6}$	&-2.65$\times10^{-6}$	&-1.93$\times10^{-6}$	&-5.44$\times10^{-7}$	&-4.44$\times10^{-7}$\\
IMSRG	&IMSRG	&-1.14$\times10^{-6}$	&-3.23$\times10^{-6}$	&-1.54$\times10^{-6}$	&-1.12$\times10^{-6}$	&-4.19$\times10^{-6}$	&-3.18$\times10^{-6}$	&5.70$\times10^{-7}$	&-4.32$\times10^{-7}$\\
    \end{tabular}
    \end{ruledtabular}
    \label{tab:4+_gs}
\end{table*}
\twocolumngrid

\end{document}